\documentclass[journal]{rmaa}
\usepackage{caption}
\usepackage{geometry}
\def\fov{FoV }
\def\RRab{RRab }
\def\RRc{RRc } 
\def\Mstd{$M_{\mbox{\scriptsize std}}$ }
\def\mins{$m_{\mbox{\scriptsize ins}}$ }
\def\fdif{$f_{\mbox{\scriptsize diff}}$ }
\def\fref{$f_{\mbox{\scriptsize dref}}$ }
\def\sref{$\sigma_{\mbox{\scriptsize ref}}$ }
\def\sdiff{$\sigma_{\mbox{\scriptsize diff}}$ }

\begin{document}

\title{Fourier Decomposition of RR Lyrae light curves and the SX Phe population 
in the central region of NGC 3201}

\author{
  A. Arellano Ferro,\altaffilmark{1,2}
  J.A. Ahumada,\altaffilmark{3}
  J.H. Calder\'on,\altaffilmark{1,3,4}
  N. Kains\altaffilmark{5}
}

\altaffiltext{1}{Visiting Astronomer, Complejo Astron\'omico El Leoncito,
operated under agreement between the Consejo Nacional
de Investigaciones Cient\'{\i}ficas y T\'ecnicas de la Rep\'ublica
\mbox{Argentina,} and the National Universities of La Plata,
C\'ordoba, and San Juan.} 
\altaffiltext{2}{Instituto de Astronom\1a, Universidad Nacional Aut\'onoma de
M\'exico, M\'exico.}

\altaffiltext{3}{Observatorio Astron\'omico, Universidad Nacional de C\'ordoba,
Laprida 854, 5000 C\'ordoba, Argentina.}

\altaffiltext{4}{Consejo Nacional de Investigaciones Cient\'ificas y T\'ecnicas -
CONICET, Argentina .}

\altaffiltext{5}{Space Telescope Science Institute, 3700 San Martin Drive, Baltimore,
MD 21218, United States of America.}

\fulladdresses{
\item A. Arellano Ferro: Instituto de Astronom\1a, Universidad Nacional Aut\'onoma de
M\'exico, Apdo. Postal 70-264, M\'exico D. F. CP 04510, 
M\'exico. (armando@astro.unam.mx)
\item J.A. Ahumada, J.H. Calder\'on: Observatorio Astron\'omico, Universidad Nacional
de C\'ordoba, Laprida 854, 5000 C\'ordoba, Argentina.
\item N. Kains: Space Telescope Science Institute, 3700 San Martin Drive, Baltimore,
MD 21218, United States of America.
}

\shortauthor{Arellano Ferro et al.}
\shorttitle{RR Lyrae stars in NGC 3201}

\SetYear{2014}
\SetVolume{50}
\ReceivedDate{May 23, 2014}
\AcceptedDate{June 9, 2014} 

\resumen{Presentamos el an\'alisis de una serie temporal de im\'agenes CCD de la
regi\'on central del c\'umulo globular NGC~3201. El objetivo principal de este
trabajo es la
descomposici\'on de Fourier de las curvas de luz de las estrellas RR~Lyrae y su
empleo en la determinaci\'on de la metalicidad del c\'umulo y de su distancia. De
esta manera hemos obtenido, para la metalicidad, el valor medio
[Fe/H]$_{ZW}= -1.483 \pm 0.006$ (estad\'istico) $\pm 0.090$
(sistem\'atico), y para la distancia, $5.000 \pm 0.001$~kpc (estad\'istico) $\pm 0.220$
(sistem\'atico). La metalicidad y la distancia estimados a partir de dos
estrellas RRc son consistentes con los anteriores. Debido a la presencia de enrojecimiento diferencial,
derivamos valores individuales de $E(B-V)$ para las estrellas RR Lyrae analizando sus
curvas de color $V-I$. El
valor promedio encontrado es $E(B-V)= 0.23 \pm 0.02$. Una exploraci\'on de las curvas
de
luz de las estrellas en la regi\'on de las {\it blue stragglers} condujo al
descubrimiento de tres nuevas variables SX Phe. La relaci\'on periodo-luminosidad de
las estrellas SX~Phe se
emple\'o para obtener una determinaci\'on independiente de la distancia al c\'umulo y
de los enrojecimientos individuales. La distancia calculada fue de 5.0 kpc.}

\abstract{CCD time-series observations of the central region of the globular cluster 
NGC~3201 were obtained with
the aim of performing the Fourier
decomposition of the light curves of the RR~Lyrae stars present in that field.
This procedure gave the mean values, for the metallicity, of 
[Fe/H]$_{ZW}=-1.483 \pm 0.006$ (statistical) $\pm 0.090$
(systematical), and for the distance,
$5.000 \pm 0.001$~kpc (statistical) $\pm 0.220$ (systematical). The values found from
two RRc stars are consistent with those derived previously. The differential reddening 
of the cluster was investigated and
individual reddenings for the RR Lyrae stars were estimated from their $V-I$ curves.
We found an average value of
$E(B-V)= 0.23 \pm 0.02$. An investigation of the light curves of stars in the {\it
blue
stragglers} region led 
to the discovery of three new SX~Phe stars. The period-luminosity relation of the 
SX~Phe stars was used
 for an independent determination of the distance to the cluster and of the
individual reddenings. We found a distance of 5.0 kpc} 


\maketitle

\section{Introduction}
\label{sec:intro}

NGC~3201 (C1015$-$461 in the IAU nomenclature) is a nearby 
($\sim$4.9~kpc) and very extended
globular cluster with a very sparse central region. These characteristics have
contributed to making
it the subject of intensive and successful searches for variable stars over
the
past one hundred years. Despite its proximity, the cluster has considerable
differential reddening (e.g., von Braun \& Mateo 2001, Piotto
et al.\ 2002; Layden \& Sarajedini 2003), which is not unexpected given its position
near the Galactic plane ($l = 277.23^{\rm o}$, $b = +8.64^{\rm o}$).
It is rich in variable stars; the 2012 update to the Catalogue of Variable Stars
in Globular Clusters (CVSGC, Clement et al.\ 2001) lists 121 variables contained
approximately within half a square degree around the cluster center, among them: 
86~RR Lyrae stars, 13~SX Phoenicis, 8~Long Period Variables, 3~eclipsing binaries, and 
 11~non-variables
previously suspected of variability. The first 56 variables in the cluster
were found
by Woods (1919) on a few Harvard plates obtained in 1916 at the Boyden Station of
Harvard Observatory near Arequipa, Peru. A few
variables, now numbered V57--V61, were added by Bailey (1922) from deep-exposure
(2- and 4-hour) plates also at Arequipa. Continuing these efforts, Dowse (1940)
announced the discovery of 25
more variables, V62--V86, on 60 new plates. Star V87 was
discovered by Wright (1941) during an
investigation of variable star periodicities, also on Arequipian plates. No
further variable search was undertaken
until 24 years later when Wilkens (1965) discovered variables V88--V96. Up to
then, the large majority of variables were of the RRab type, four were RRc, and a few
others would turn out to be non-variable in later more precise studies. Stars
V97--V100 were found from photometric and photographic data by Lee (1977). More
than twenty years later, already in the CCD era, von Braun \& Mateo (2002) found the
eclipsing binary V101 and other short-period variables in the field of the cluster,
mostly eclipsing binaries and one RR~Lyrae which they argued were not cluster
members. The first group of SX~Phe stars, V102--V112, was identified by Mazur et al.\
(2003). Layden \& Sarajedini (2003) (LS03) detected low-light variations in
some bright red
giants (V113--V118), 
and discovered another eclipsing binary (V119) and two  probable SX~Phe stars
(V120 and V121). 

The present investigation is the first one that uses the difference image analysis
(DIA) technique on NGC 3201.
The virtues of DIA as a powerful tool to discover short-period
variable stars, or unveil small amplitude variations in Blazhko RR Lyrae stars 
in the densely populated 
central regions of globular clusters, have been
shown in several papers in the recent literature (e.g. Arellano Ferro et al.\
2013a, 2012; Bramich et al.\ 2011; Figuera Jaimes et al.\ 2013; 
Kains et al.\ 2013, 2012 and references 
therein). We applied this method to the
central regions of NGC 3201 on a set of data obtained under average-to-mediocre
seeing conditions; in spite of this, our photometry led to clean light curves for
the
great majority of the RR Lyrae stars and allowed
the discovery of three new SX Phe as shall be described later. 

The principal aim of the present paper is to offer a new time-series \emph{V} and
\emph{I} CCD
photometry that
allows the refining of the periods of the variables as well as the Fourier
decomposition of the light
curves of the RR~Lyrae population to estimate its iron content [Fe/H]
and distance independently; we also want to compare the cluster distance
obtained via
the SX~Phe Period-Luminosity (P-L)
relation to estimates derived with other methods. In the process we shall discuss the
peculiarities of several stars of the
cluster population of variables.

The layout of the paper is as follows: In Sect.~2 we describe the observations,
data reductions, and transformation of our photometry to the 
standard system. In Sect.~3 the distribution of the variable stars
in the Color-Magnitude diagram (CMD) is presented and three new SX~Phe stars are
announced. The RR~Lyrae are identified, their periods are refined, and their light
curves are presented.
In Sect.~4 the individual reddenings of RR Lyrae stars are calculated
and the Fourier decomposition of their light curves is
performed to estimate their metallicity, luminosity, mass, and radius. The
distribution
of the RR~Lyrae stars on the Bailey diagram (Amplitude versus Period) is discussed. In
Sect.~5 we address the P-L relation for SX~Phe stars and comment on the
implied cluster distance. Sect.~6 summarizes our results.

\section{Observations and reductions}
\label{sec:ObserRed}

\subsection{Observations}

The Johnson-Kron-Cousins $V$ and $I$ observations 
used in the present work were obtained on  March 19--22, 2013, with
 the 2.15-m telescope of the Complejo Astron\'omico El Leoncito (CASLEO), 
 San~Juan, Argentina. The estimated seeing varied between $\sim 1.5$ and 2.8~arcsec, with
a tendency to increase as each night progressed.
The detector was a Roper Scientific
back-illuminated CCD of $2048 \times 2048$ pixels with a 
scale of 0.15~arcsec/pix and a field of view (FoV) 
of approximately $5.1 \times 5.1$~arcmin$^2$. Our data consist of 143~$V$ and 146~$I$
images.

\subsection{Difference Image Analysis (DIA)}
\label{DIA}

Image data were calibrated using bias and flat-field
correction procedures. We used DIA
to extract high-precision time-series photometry in the field of NGC~3201. As in
previous papers, we used the 
{\tt DanDIA}\footnote{{\tt DanDIA} is built from the DanIDL library of IDL routines
available at \texttt{http://www.danidl.co.uk}}
pipeline for the data reduction process (Bramich et al.\ 2013), which includes an 
algorithm that models the convolution kernel matching the PSF
of a pair of images of the same field as a discrete pixel array (Bramich 2008). 

A brief summary of the DanDIA pipeline can be found e.g., in the paper by Arellano
Ferro et al.\ (2013a),
while a detailed description of the procedure and its caveats are available in 
 the paper by
 Bramich et al.\ (2011), to which the interested reader is referred for 
the relevant details.

\begin{figure}[!t]
\begin{center}
\includegraphics[width=8.0cm,height=8.0cm]{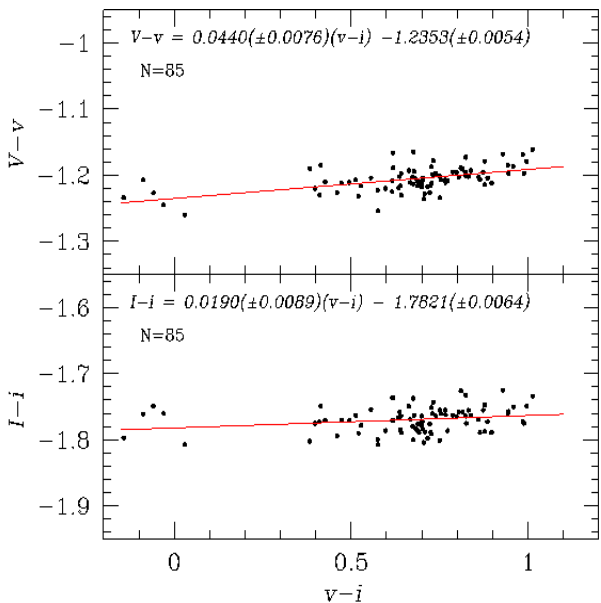}
\caption{The transformation relationship between the instrumental and standard
photometric systems using a set of standards of Stetson (2000) in the FoV of our images
of NGC~3201.}
    \label{trans}
\end{center}
\end{figure}

We also used the methodology developed by
Bramich \& Freudling (2012) to solve for the 
magnitude offset that may be introduced into the photometry 
by the error in the fitted value of the photometric scale factor
corresponding to each image.
The magnitude offset due to this error can reach up to 
$\approx30$~mmag in a few cases, but it is generally of the order
of $\approx1{-}10$~mmag. 
This correction improved the quality of the light curves,
particularly for the brightest stars.

\subsection{Transformation to the \textit{VI} standard system}

From the high number of standard stars of Stetson (2000)\footnote{%
 \texttt{http://www3.cadc-ccda.hia-iha.nrc-cnrc.gc.ca/\\
community/STETSON/standards}}
in the large field of NGC~3201, 85 were identified in the
FoV of our images with $V$ and $V-I$ in the ranges 12.6--18.6~mag and 
$-0.14$--$1.01$~mag,
respectively. These were used to transform our instrumental system into the
Johnson-Kron-Cousins photometric system (Landolt 1992). The standard minus the
instrumental magnitude differences show a mild dependence on the color as displayed
in Fig.~\ref{trans}. The transformation equations are of the form:
 %
%
%
\begin{eqnarray}\label{eq:transV}
V &=& v +0.0440\, (\pm0.0076)\, (v-i) \nonumber \\
  && -1.2353\, (\pm0.0054),
\end{eqnarray}
\begin{eqnarray}\label{eq:transI}
I &=& i +0.0190\, (\pm0.0089)\, (v-i) \nonumber \\
  && -1.7821\, (\pm0.0064).
\end{eqnarray}

\section{Variable Stars in NGC 3201}

The variable stars in our FoV are listed in Table~\ref {variables}
along with their
mean magnitudes, amplitudes, and periods derived from our photometry. The coordinates
listed in columns~10 and~11 are taken from the CVSGC, and were calculated by Samus et
al. (2009) and Mazur et al. (2003). We found them to be
accurate and no attempt to recalculate them was made. We include them here for
 the sake of completeness.

\begin{figure}
\begin{center}
\includegraphics[width=8.0cm,height=8.0cm]{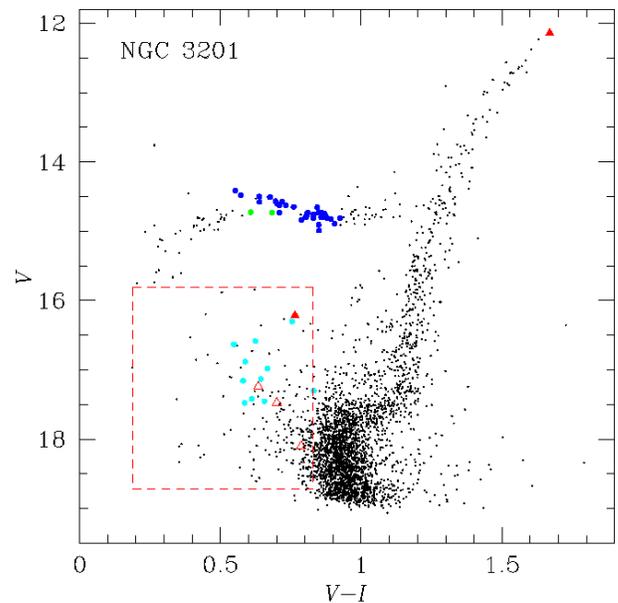}
\caption{CMD of NGC 3201. The magnitudes and colors plotted are magnitude-weighted
means over our entire collection of images. Blue and green circles are,
respectively, RRab and RRc variables in our FoV. An arbitrarily defined
\textit{blue~stragglers} region is enclosed by the dashed lines. Known SX~Phe stars are
represented by cyan circles. We searched for light variability in all stars
inside that region and discovered three new SX Phe shown as red open triangles.
The SR variable V117 and the eclipsing binary V119
are represented by solid red triangles.}
\label{CMD}
\end{center}
\end{figure}

\begin{figure*}
\begin{center}
\includegraphics[width=15.cm,height=15.cm]{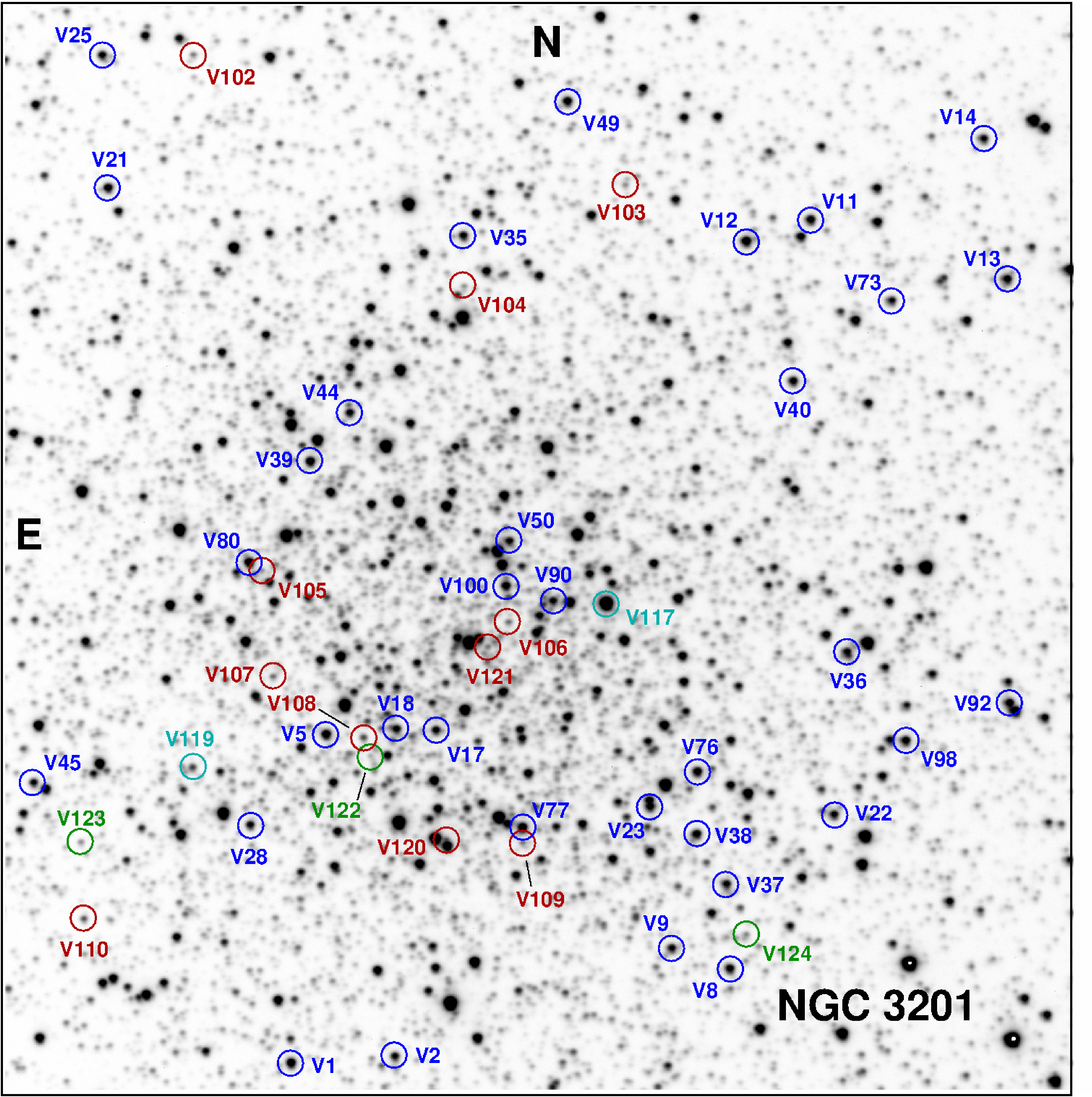}
\caption{Finding chart constructed from 9 of our best images in $V$. The field
is about $5.1\times5.1$~arcmin$^{2}$. Labels are: blue for known RR~Lyrae stars,
red for known SX~Phe stars, and green for the three SX~Phe discovered in this work.
The SR~variable V117 and the eclipsing binary~V119 are labeled in light blue.}
\label{chart}
\end{center}
\end{figure*}

\begin{table*}
\scriptsize
\begin{center}

\caption{General data for all of the confirmed variables in NGC~3201 in the \fov of 
our images.  Previous period estimates for each variable from LS03 are reported in
column~7 for comparison with our refined periods in
column~9. The period uncertainties in parenthesis correspond to the last decimal
places.}
\label{variables}

\begin{tabular}{llllllllllll}
\hline
Variable & Variable & $<V>$ & $<I>$   & $A_V$  & $A_I$   & $P$ (LS03) &  HJD$_{\mathrm{max}}$ 
& $P$ (this work)    & RA   & Dec         \\
Star ID  & Type     & (mag) & (mag)   & (mag)  & (mag)   & (d)  & 
($+2\,450\,000$) & (d) & (J2000.0)   & (J2000.0)    \\
&&&&&&&&&& & \\
\hline
V1       & RRab     & 14.841  & 13.887  & 0.951       &0.560  & 0.6048761       
       & 6373.5087      & 0.604811(4)     & 10:17:42.82 & $-$46:26:37.8 \\     
V2       & RRab     & 14.839  & 14.002  &  1.081   & 0.653   &0.5326722         
        & 6373.5922    & 0.532621(5)      & 10:17:39.98 & $-$46:26:37.9\\
V5       & RRab      & -- & --  &1.231     & --  & 0.511550       
        & 6373.4929       & 0.501345(4)      & 10:17:41.28 & $-$46:25:06.3\\    
V8       & RRab     &14.760   & 13.884 & 0.475       & 0.303   & 0.6286573     
       & 6371.7232     & 0.628568(5)   & 10:17:30.75 & $-$46:26:19.3\\   
V9       & RRab      &14.834   & 14.034  &0.820   & 0.485  & 0.525530       
       & 6374.7383       & 0.525397(4)      & 10:17:32.28& $-$46:26:12.6\\
V11       & RRc     &14.808   &14.136 &  0.49  & 0.312   & 0.299049         
        & 6373.5632     & 0.299134(3)   & 10:17:27.55& $-$46:22:55.6\\
V12       & RRab     &--  & --  & --    & --  & 0.4955547     
        & 6373.5087      &0.497369(4)    & 10:17:29.03 & $-$46:22:55.8\\       
V13       & RRab     & 14.883  &14.070  & 0.894  & 0.603   & 0.5752145       
       & 6371.7280      & 0.574822(5)  & 10:17:22.03 & $-$46:23:11.2\\
V14       & RRab      & 15.009  &-- &1.130  &  --  & 0.5092945         
       & 1216.7563     & 0.508941(5)     & 10:17:22.42 & $-$46:22:31.5\\      
V17      & RRab    & 14.820 &13.960 & 0.840  & 0.566   & 0.565590     
       & 6371.5406       & 0.565844(5)   & 10:17:38.29 & $-$46:25:07.2\\
V18       & RRab Bl:    & 14.812  & 13.983  & 0.732    & 0.494   & 0.539655      
       & 6373.7508       & 0.540442(5)      & 10:17:39.40  & $-$46:25:07.0\\
V21       & RRab     & 14.858 & 13.979  &  0.723   & 0.446   &0.566628        
        & 6371.7429      & 0.566754(5)      & 10:17:46.23 & $-$46:22:29.4\\
V22       & RRab      & 14.753 & 13.897  & 0.833    & 0.503   & 0.6059882       
        & 6373.7086       & 0.605843(5)     & 10:17:27.64 & $-$46:25:38.0\\   
V23       & RRab     &14.818   &13.951 & 0.771  & 0.501   & 0.586775      
       & 6372.7320     & 0.586776(5)   & 10:17:32.64 & $-$46:25:31.1\\    
V25       & RRab Bl:     &14.787   & 14.029   &0.633   & 0.468  & 0.514804       
       & 6373.6127     & 0.514696(3)       & 10:17:46.15 & $-$46:21:52.1\\ 
V28       & RRab Bl:     &14.885    & 13.981  &  0.981  & 0.57--0.77  & 0.580025     
        & 6373.6570     & 0.579506(5)   & 10:17:43.50 & $-$46:25:30.4\\
V35       & RRab     &14.771  & 13.912  & 0.619    & 0.400  & 0.6155244      
        & 6371.5674      &0.615523(5)   & 10:17:36.29 & $-$46:22:43.2\\       
V36       & RRab     & 14.751   & 14.051  &0.655  & 0.341  &0.484178      
       & 6373.6570      & 0.479576(5)   & 10:17:27.01 & $-$46:24:53.0\\
V37       & RRab      & 14.785  &13.932 & 0.746  & 0.476 & 0.575740         
       & 6372.6919     &0.575145(4)     & 10:17:30.69 & $-$46:25:55.7\\      
V38      & RRab    & 14.836 &-- &0.734   & --   & 0.509090        
       & 6373.4900      &0.509405(5)  &10:17:31.42 & $-$46:25:41.1\\
V39       & RRab     & 14.879 &-- & 1.244  & --  & 0.483233      
       & 6373.4900       &0.482872(5)   & 10:17:41.23 & $-$46:23:49.7\\       
V40       & RRab     & 14.854 & 13.948  &  0.403   & 0.279   &0.643820        
        & 6373.6100     & 0.643744(4)     &10:17:28.00 & $-$46:23:35.9\\
V44       & RRab      & 14.803&13.933  &  0.687    & 0.411 & 0.6107344        
        & 6373.6991       &0.610663(4)      &10:17:40.08 & $-$46:23:36.7\\     
V45       & RRab     &14.974   & 14.089  & 1.002 & 0.629   & 0.537460      
       & 6373.6127     & 0.538058(4)   &10:17:49.29 & $-$46:25:14.6\\   
V49       & RRab      &14.760   & 13.946   &0.896   & 0.586  &0.581020      
       &6371.6775      &0.580957(4)      & 10:17:33.63 & $-$46:22:13.4\\ 
V50       & RRab Bl:    &14.832   & 14.020  &  0.916 & 0.582  & 0.542178         
        & 6373.6745    & 0.542347(5)    & 10:17:36.10  & $-$46:24:14.0\\
V73       & RRab Bl    &14.816   & 14.061 & 1.06--1.38  & 0.740  & 0.519965      
        & 6373.5991      &0.519538(5)    & 10:17:25.20 & $-$46:23:15.2\\       
V76       & RRab     & 14.821 & -- &0.623   & --  & 0.526680       
       & 6373.4900      &0.526736(4)  & 10:17:31.29 & $-$46:25:23.6\\
V77       & RRab      & 14.693   &13.891  & 1.079  & 0.529  & 0.567644         
       & 6373.5303     &0.567393(7)     &10:17:36.13 & $-$46:25:36.1\\      
V80      & RRab    & 14.650  & 13.910 & 0.810 & 0.618   & 0.589960        
       & 6373.5108    & 0.588706(7)    & 10:17:43.07 & $-$46:24:16.8\\
V90       & RRab     & 14.753   & 13.935 & 1.081  &0.58--0.66   & 0.606105      
       & 6373.6680     & 0.606387(6)      & 10:17:34.95 & $-$46:24:33.3\\
V92       & RRab     & 14.736 & 13.934  & 1.083    & 0.557   &0.539585         
        & 6373.5534     &0.539528(6)     & 10:17:22.70  & $-$46:25:11.0\\
V98       & RRc      & 14.759 & 14.053  &  0.38--0.46 & 0.243   & 0.335647        
        & 6373.5322   &0.336259(5)      & 10:17:25.14 & $-$46:25:20.0\\  
V100       & RRab     &14.813 & -- &  0.986   & --   & 0.548920     
       & 6373.4900     &0.548513(6)   &10:17:36.14 & $-$46:24:27.9\\ 
V102       & SX Phe      &17.164    & 16.574   &0.060   &    & --       
       & 6374.7421   & 0.045398(18)     &10:17:43.50 & $-$46:21:53.4\\   
V103       & SX Phe      &17.424   & 16.804   &0.054   & --  & --       
       & 6373.6570   & 0.037241(62)  & 10:17:32.10 & $-$46:22:37.9 \\ 
V104       & SX Phe     &17.457   &16.800   & 0.038   & --   & --         
        &6371.7233   & 0.037510(73)    & 10:17:36.60 & $-$46:23:03.4\\
V105       & SX Phe     &17.506  & 16.894  &  0.060   & --  &--      
        & 6374.7621    &0.037496(10)    & 10:17:42.60 & $-$46:24:19.5\\ 
V106       & SX Phe     & 16.883  & 16.291 & 0.034  & --  & --       
       & 6371.6818     & 0.043552(13)  &10:17:36.00 & $-$46:24:38.2\\
V107       & SX Phe      & 16.982  & 16.307& 0.126 &  --  & --         
       & 6374.7503  & 0.040476(97)    &10:17:42.50& $-$46:24:49.1\\
V108      & SX Phe    & 16.325&15.551 &0.236   & --   & --        
       & 6373.6897    & 0.067349(55)    &10:17:40.20 & $-$46:25:07.8\\
V109       & SX Phe      &17.176   &16.498 & 0.302  & --  & --       
       & 6373.5394       & 0.054259(41)     & 10:17:36.00 & $-$46:25:40.2\\
V110       & SX Phe     &16.603   & 15.965 & 0.249  & --  & --         
        & 6372.5672     & 0.050137(19)    &10:17:48.10 & $-$46:25:53.4 \\
V117       & SR     & -- &  --  & --   & --  & --      
        & --      &--    & 10:17:33.45 & $-$46:24:34.6\\
     
V119       & EA     & 16.218  & 15.45 & 0.04  & 0.04  & --       
       & 6372.6511  &1.2759 & 10:17:44.98 & $-$46:25:13.3\\
V120       & SX Phe      & 16.677  & 16.114  & 0.367 &--  & --       
       & 6372.6006     & 0.088162(46)     & 10:17:38.30 & $-$46:25:38.0\\ 
V121      & SX Phe    &17.312 & 16.475& 0.127  & --  & --        
       & 6373.5602    & 0.037369(18)   & 10:17:36.58 & $-$46:24:43.2\\

V122$^{a}$       & SX Phe     & 18.107  & 17.324 &0.053   & --  & --    
       & 6371.7700      & 0.067279(48) & 10:17:40.15 & $-$46:25:12.7  \\
V123$^{a}$       & SX Phe      & 17.480  &16.804 &0.058   & --& --       
       & 6374.7657     & 0.034895(8)     &10:17:48.12 & $-$46:25:32.2  \\      
V124$^{a}$     & SX Phe    & 17.245&16.607 & 0.022  & -- & --       
       & 6372.6720    & 0.040128(32)   &10:17:30.22& $-$46:26:09.1 \\
\hline
\end{tabular}
\raggedright
\center{\quad $^{a}$newly found in this work.}
\end{center}
\end{table*}

\subsection{The Color-Magnitude diagram}

The CMD of the cluster is shown in Fig.~\ref{CMD}, where the location of the
known and the newly discovered 
variables is marked. Individual stars are discussed in the following
section.
The red-dashed area is an arbitrarily defined \textit{blue stragglers} region were SX~Phe
variables are expected to be found. Indeed, examination of the light curves of all
stars in this
region confirmed 
the nature of all previously known SX~Phe stars and led to the discovery of three
more
(see \S~\ref{Sec:SXPHE}).  The finding chart of all variables in Table~\ref{variables} 
is in Fig.~\ref{chart}. 
 Given the differential nature of the reddening in front
of the cluster we made no attempt to deredden the CMD. An interesting and
successful exercise of this procedure may be found in LS03.

\subsection{RR Lyrae stars}
\label{sec:RRL}

\begin{figure*}
\includegraphics[width=16.cm,height=16.cm]{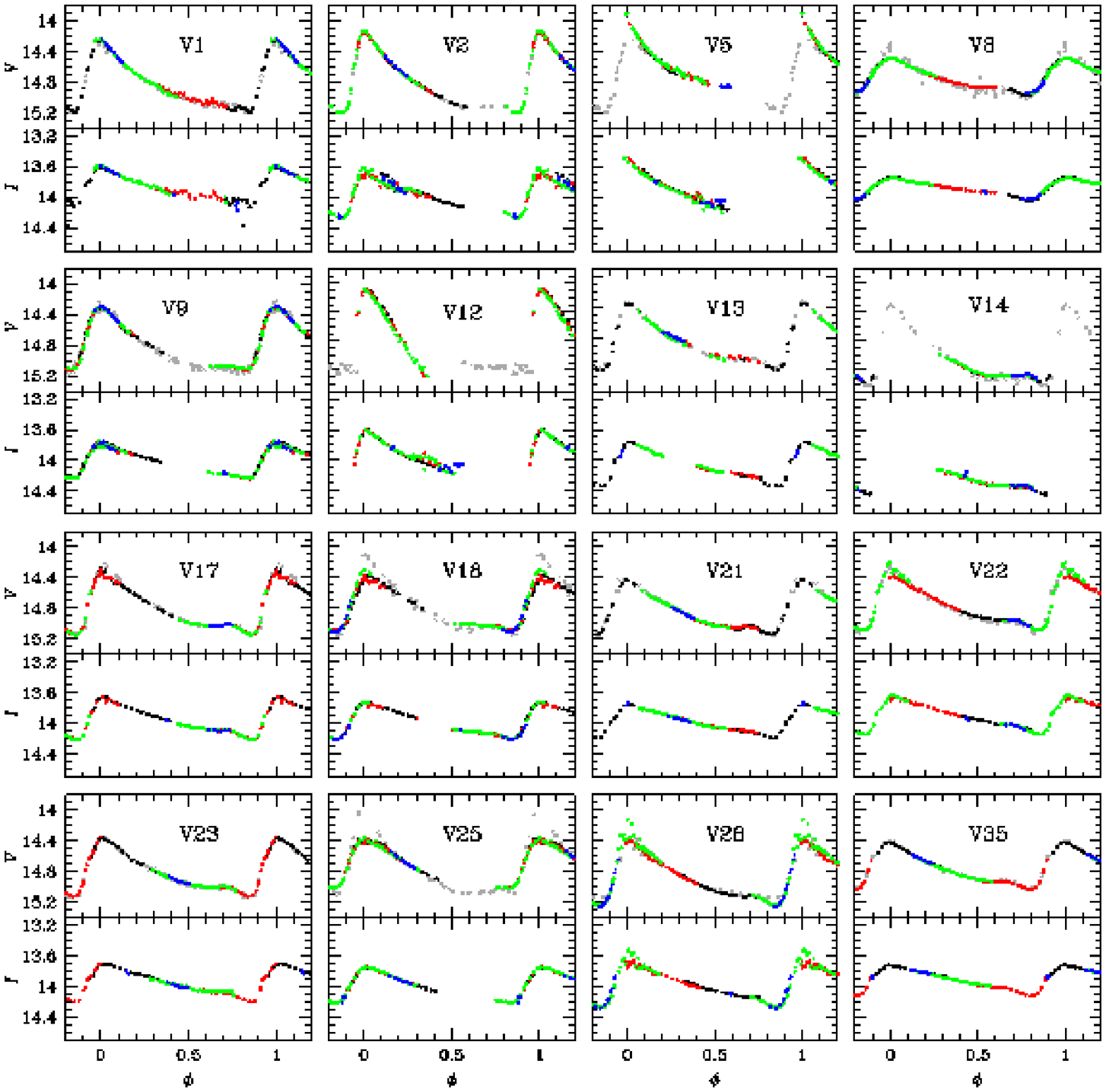}
\caption{Light curves of the RRab stars in our FoV
phased with the periods listed in Table~\ref{variables}. The vertical scale is the
same for all curves. The symbols in black, red, green, and blue
correspond to each one of the four nights of our run, respectively. Gray symbols are data from LS03 and are
included as a
reference. See text for discussion on individual stars.}
    \label{RRL_A}
\end{figure*}

A comparison of the intensity-weighted means $<V>$ and $<I>$ (Table \ref {variables})
and those of LS03 for 27 RR Lyrae stars in common indicates
that, on average, our $<V>$ values are 0.02 mag fainter and the $<I>$ values are 0.004~mag
brighter, i.e., well within photometric errors. Therefore, the
periods of the RR Lyrae stars were calculated by combining the $V$ data from
LS03 obtained in 1999, with our $V$ photometry using the
string-length method (Burke, et al.\ 1970; Dworetsky 1983). 
The 14 year
time-base allows a precise determination of the periods. In Table~\ref{variables}
the new periods and epochs are listed, and for comparison the periods of LS03 
are included as well. The $V$ and $I$ light curves of RRab and RRc stars are shown in 
Figs.~\ref{RRL_A} and \ref{RRLc} respectively, having been phased by 
using our refined periods and the epochs listed in Table~1. In the
light curves, observations from different nights are plotted 
with different colors for clarity; data from LS03 are also plotted in light gray.
 We can see that the LS03 light curves and ours match very well
and are properly phased with the new periods. Later in this paper, when dealing
with the Fourier
decomposition, we shall exclusively use our data except in a few
cases where data from LS03
nicely complete a light curve not properly covered otherwise, e.g., V14, V38, or V39.
Plotting both data sets unveils long-term amplitude modulations of the Blazhko type
not easily evident in our four-night light curves. These probable Blazhko variables,
V18, V25, V28, V50, and V73 are labeled ``Bl'' or ``Bl:'' in
Table \ref {variables}.

\begin{figure*} 
\includegraphics[width=16.cm,height=16.cm]{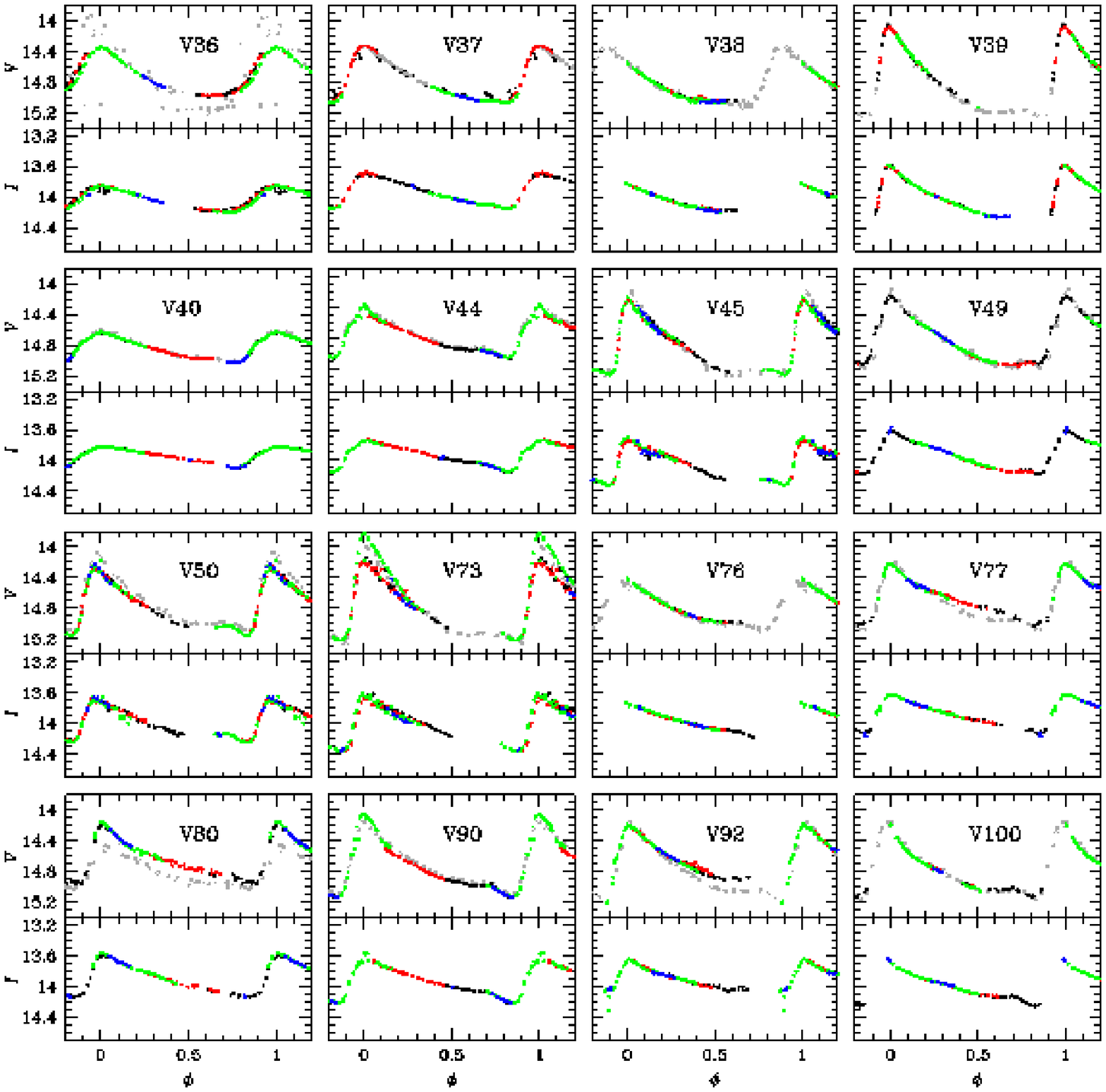}
\caption{Fig 4. Continued}
    \label{RRL_B}
\end{figure*}

\subsection{SX Phe stars}
\label{Sec:SXPHE}

Since our photometry spans only four nights and the seeing was not particularly good,
we refrained from performing the statistical approach we followed in previous
papers (e.g., Arellano Ferro et al.\ 2013a,b; Figuera Jaimes et al.\ 2013) to identify
new variable
stars in our
light curve collection. Instead, we examined the light curves of all stars lying on
 the blue straggler region shown in the CMD  of Fig.~\ref{CMD}. This procedure
allowed us to recover all known or
suspected SX~Phe stars in the CVSGC, as well as to find three new ones, which we labeled
V122, V123, and V124. These stars will be discussed in detail later in this paper.
The light curves of the 14 currently known SX~Phe stars in the cluster are shown in
Fig.~\ref{SXphe} and were phased with main periods calculated with {\sf
period04} (Lenz \& Breger 2005). The ephemerides and amplitudes of the main frequencies
are listed in Table~\ref{variables}. 

The appearance of many of the light curves strongly suggests the presence of multiple
excited
modes. In fact, as reported in the discovering paper (Mazur et al.\ 2003), several frequency modes
can be identified in the majority of these SX~Phe stars (see their
Table~3) and therefore we were motivated to carry out an independent search of active modes in them.
Although our photometry did not permit the detection of the many secondary frequencies
reported by Mazur et al.\ (2003), we could confirm the principal frequencies and, 
in some stars, we were able to find secondary frequencies which led to 
the identification of the modes. Our frequency findings are given in Table~\ref{SXfreq}.

Variables V120 and V121 are labeled as ``SXP?'' in the 2012 CVSGC. This classification
comes most likely from Table~8 of LS03,
where these stars (776 and 941 respectively) are labeled as ``$\delta$ Scuti?''. Given
their frequencies in Table~\ref{SXfreq} and their position in the blue stragglers region,
we confirm the SX~Phe character of both stars. We
also performed a frequency
analysis of the new SX~Phe stars V122, V123, and V124. In stars V103, V106, 
and V108 two frequencies were detected and identified with the radial
fundamental 
and first overtone modes, exhibiting ratios $f_1/f_2$ very
close to the predicted value of 0.783 (see Santolamazza et al.\ 2001; Jeon et al.\
2003; Poretti et al.\ 2005). One of the pulsating modes of V107  seems to be
non-radial.

\begin{figure}
\includegraphics[width=8.cm,height=5.cm]{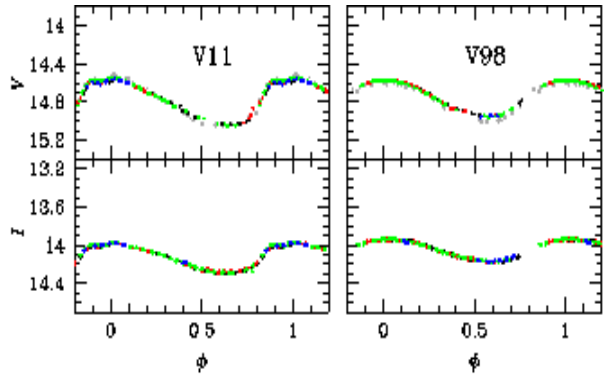}
\caption{Light curves of the two RRc stars in our FoV. Symbols are as in 
 Fig.~\ref{RRL_A}.}
    \label{RRLc}
\end{figure}

\begin{table*}
\scriptsize
\begin{center}
\caption{Frequencies and amplitudes of the SX~Phe stars in NGC~3201. The
numbers in parentheses indicate the uncertainty on the last decimal places.}
\label{SXfreq}
\begin{tabular}{llllllll}
\hline
Variable &$V$& $V-I$& $E(B-V)^{a}$&$f$& $A$& $P$& Mode id.$^{b}$\\
& & && (d$^{-1}$) & (mag) & (d)& \\
\hline
V102& 17.164&0.581&0.30&22.027&0.060& 0.045398(18)&$f_1$:F\\
V103& 17.423&0.611&0.23&26.852&0.054& 0.037241(11)& $f_1$:F\\
    &       &     &&34.367&0.045& 0.029097(8)& $f_2$:1O; $f_1/f_2$=0.781\\
V104& 17.457&0.656&0.23&26.659&0.038& 0.037510(43)& F\\
V105& 17.506&0.586&0.23&26.670&0.060& 0.037496(10)&F \\
V106& 16.883&0.588&0.11&22.961&0.034&0.043552(13)&$f_1$:F\\
    &       &     &&29.240&0.023&0.034200(39)&$f_2$:1O; $f_1/f_2$=0.785\\
V107& 16.982&0.667&0.23&24.706&0.126&0.040476(97)&$f_1$:F\\
    &       &     &&25.245&0.081&0.039610(33)&$f_2$:non-radial\\
V108& 16.325&0.754&0.11&14.848&0.236&0.067349(55)&$f_1$:F\\
    &       &     &&19.073&0.122&0.052430(92)&$f_2$:1O; $f_1/f_2$=0.778\\
V109& 17.176&0.643&0.29&18.430&0.302&0.054259(41)&$f_1$:F\\
V110& 16.603&0.624&0.23&19.945&0.249&0.050137(19)&$f_1$:F\\
V120& 16.677&0.548&0.33&11.342&0.367&0.088162(46)&$f_1$:F\\
V121& 17.312&0.832&0.23&26.760&0.127&0.037369(18)&$f_1$:F\\
V122& 18.107&0.787&--&14.863&0.053&0.067279(48)&$f_1$:F\\
V123& 17.480&0.700&0.23&28.658&0.058&0.034895(8)&$f_1$:F\\
V124& 17.245&0.634&0.23&24.920&0.022&0.040128(32)&$f_1$:F\\
\hline
\end{tabular}
\center{\quad $^{a}$See $\S$~\ref{sec:SXP} for a discussion on the individual
reddenings; $^{b}$F: Fundamental mode; 1O: First overtone }
\end{center}
\end{table*}

\begin{figure*} 
\includegraphics[width=14.cm,height=14.cm]{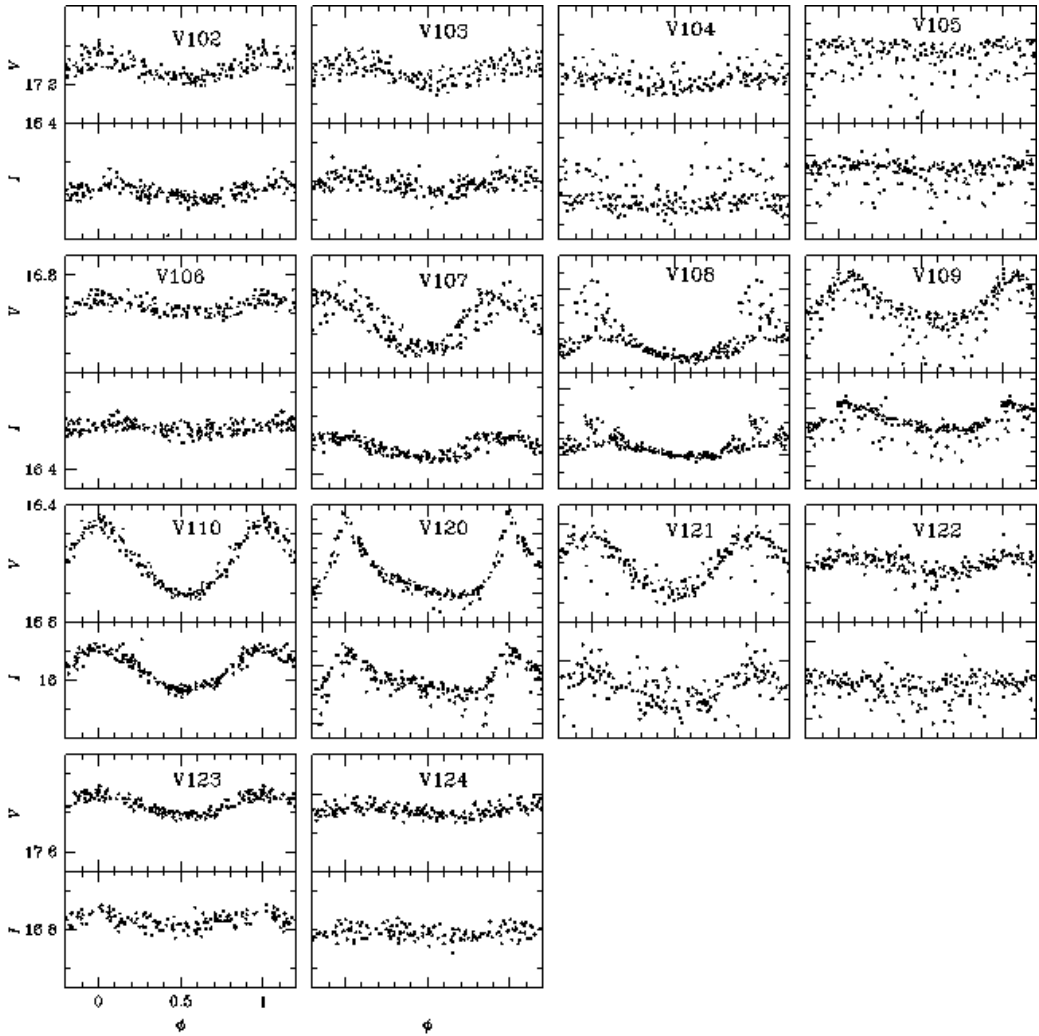}
\caption{Light curves of the SX~Phe stars in our FoV
phased with the main period listed in Table~\ref{variables}. The vertical scale
is not the same for all curves; the tick mark on the vertical axis is equivalent to 
0.1~mag. Variables V122, V123, and V124 are new discoveries.}
    \label{SXphe}
\end{figure*}

\subsection{Comments on individual variables}
\label{sec:indvar}

In this section we only discuss those variable stars  that
 deserve particular comments.

{\bf V5, V11, V12, V13, V39, V76, V77, V80, V98, V100.}
As mentioned before, in the reported observations the seeing appears to
worsen along the night. This
led to additional scatter in the light curves of these stars due to 
contamination by neighbouring stars, particularly near the minima. We thus either
neglected as much as possible or did not include them in
the further analysis (e.g., stars V5 and V12).

{\bf V14.} Our phase coverage for this light curve is incomplete, and as a result,
our values of $<V>$ and $<I>$ are biased. We therefore used the values from LS03 to
plot the 
star in the CMD and included LS03 data to complete the light curve 
in order to perform Fourier decomposition.

{\bf V25 and V36.} The shapes of the light curves of these stars are peculiar, showing 
 very small amplitude for their period. In the case of V25, the addition of the LS03
light curve
hints at the presence of amplitude modulations. For star V36,
LS03 data are very scattered and do
not follow our light curve. These stars will stand out as peculiar in the Bailey
diagram discussed later (\S~\ref{sec:Bailey}).

\begin{figure}
\includegraphics[width=8.cm,height=7.cm]{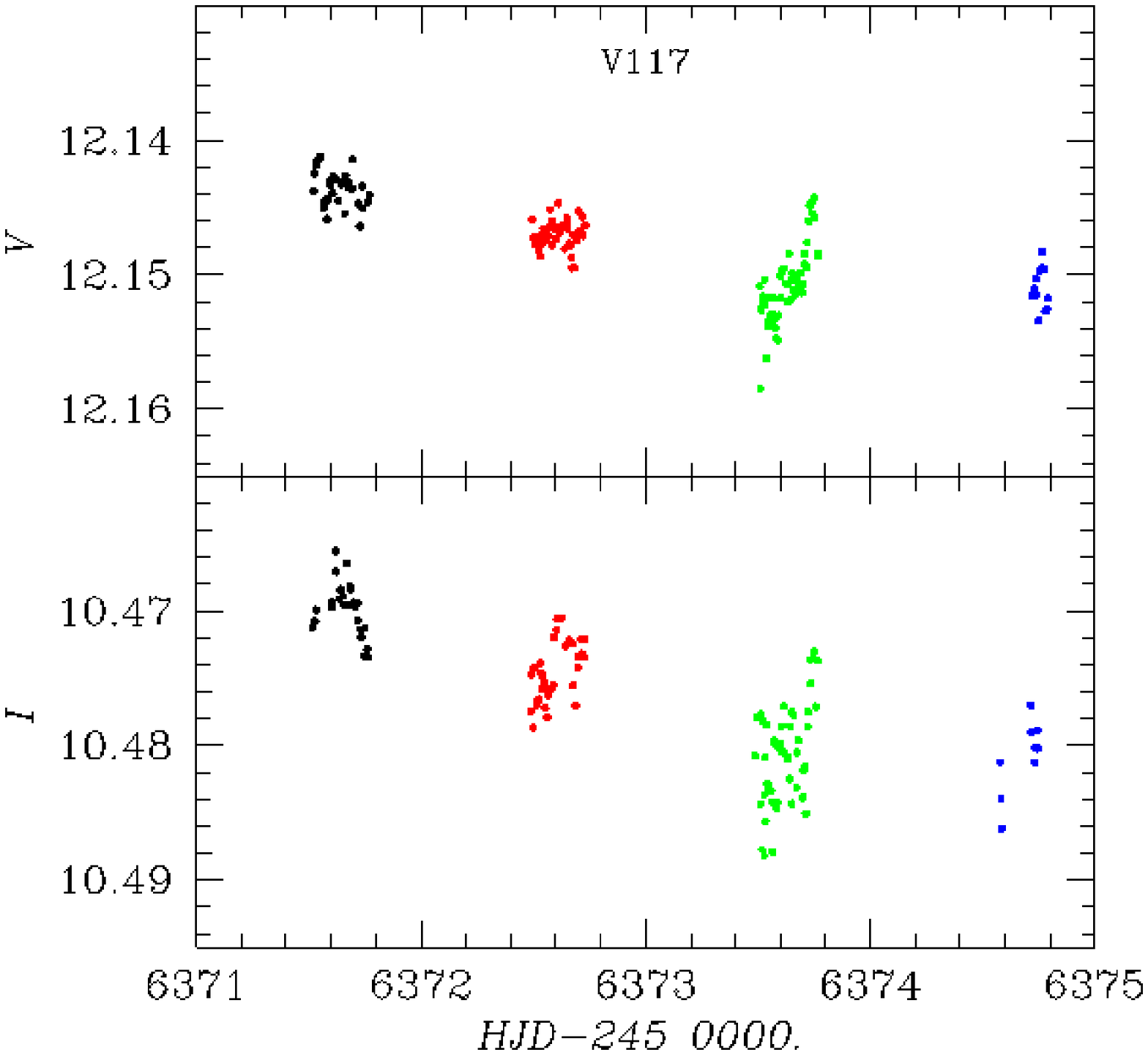}
\caption{$V$ and $I$ light curves of the SR variable V117. 
Intra-night variations are likely due to
noise from seeing variations. Gradual fainting over the four nights  should be real.}
    \label{V117}
\end{figure}

{\bf V117.} The light curve of this star, which is the
brightest of the cluster in our FoV, noticeably degraded after the post-calibration
process described in \S~\ref{DIA}, which seems to have failed
in this particular case. Hence, the light curve shown in Fig.~\ref{V117} 
was not post-calibrated. 
The intra-night variations are likely not real but originated in the
seeing variations. The real variation is reflected as a gradual dimming over 
the four nights.

\begin{figure}
\includegraphics[width=8.cm,height=7.cm]{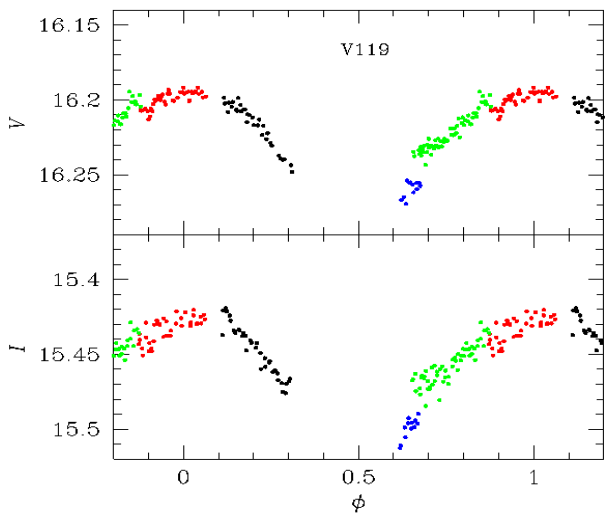}
\caption{$V$ and $I$ light curves of the eclipsing binary V119 phased with $P=1.2759$~days.}
    \label{V119}
\end{figure}

{\bf V119.} Fig.~\ref{V119} shows the light curve of this eclipsing binary phased
with a period of 1.2759~days. With our data it is not possible to distinguish between
this period and another of 0.595941~days.

\section{Physical parameters of RR~Lyrae stars}
\label{sec:Four}

The shape of the light curve of a RR~Lyrae star carries information of some of its
physical parameters of astrophysical relevance such as [Fe/H], log$(L/{\rm L_{\odot}})$, and
log~$T_{\rm eff}$. These can be calculated by means of the Fourier decomposition of the light curve,
which is performed by fitting it with a series model of the form: 
\begin{equation}
\label{eq.Foufit}
m(t) = A_0 + \sum_{k=1}^{N}{A_k \cos\ ({2\pi \over P}~k~(t-E) + \phi_k) },
\end{equation}
%
where $m(t)$ is the magnitude at time $t$, $P$ is the period, and $E$ is the epoch. A
linear
minimization routine is used to derive the best fit values of the 
amplitudes $A_k$ and phases $\phi_k$ of the sinusoidal components. 
From the amplitudes and phases of the harmonics in Eq.~\ref{eq.Foufit}, the 
Fourier parameters, defined as $\phi_{ij} = j\phi_{i} - i\phi_{j}$, and $R_{ij} = A_{i}/A_{j}$, 
are computed. 

It has been shown that $ad~hoc$ semi-empirical calibrations can correlate the above
Fourier parameters 
with the physical quantities of interest. Although numerous calibrations exist in the
literature, in previous papers we have 
argued in favour of those developed by Jurcsik \&
Kov\'acs (1996) and
Kov\'acs \& Walker (2001) for the iron abundance and absolute magnitude of RRab 
stars, and those of
Morgan, Wahl \& Wieckhorts (2007) and Kov\'acs (1998) for  RRc stars. 
The effective temperature $T_{\rm eff}$ was estimated using the calibration of Jurcsik
(1998). These calibrations and 
their zero points have been discussed in detail in Arellano Ferro et al.\ (2013a).

Due to the presence of differential reddening, before estimating the distance to the
cluster from its RR~Lyrae stars it is necessary to individually correct their
magnitudes. We estimated the individual color excesses using the method originally
proposed by
Sturch (1966), and further investigated
by Blanco (1992), Mateo et al.\ (1995), and Guldenschuh et al.\ (2005).
Guldenschuh et al.\ (2005) concluded that, for RRab stars, the intrinsic color
between phases 0.5 and 0.8 is $(V-I)_0^{\phi(0.5-0.8)}=0.58 \pm 0.02$~mag.
We have already successfully applied this method to the 
RR~Lyrae stars of another cluster with heavy differential reddening, NGC 6333 (Arellano
Ferro et al.\ 2013a). 
In the case of the present work, however, we noted that in several
of the RRab light curves in Fig.~\ref{RRL_A} the
bump appears in phases smaller than 0.8, so  we restricted the calculation of $<V>$
and $<I>$ to the 0.5--0.7 range.
Then we calculated $E(B-V)=E(V-I)/1.616$. The resulting reddenings for the RRab stars
are
listed in column~10 of Table~\ref{tab:fourier_coeffs}. Values of $E(B-V)$ for stars
with scanty or peculiar data in the above phase range  (e.g., V50,
V77, and V80) are marked with a colon, and for them we instead adopted the mean reddening 
value in subsequent calculations.
The same approach was followed by LS03 and their values are
listed in their Table~6. A comparison of both sets of reddenings for the 14 stars in
common reveals that, with a scatter of 0.03~mag, our values are on average 0.03~mag
smaller.  The average of our reddenings is $0.23 \pm 0.02$, which compares well
with the value $0.25 \pm 0.02$ of LS03 for stars within 2~arcmin of the cluster
center.

The value of $A_0$ and the Fourier light-curve fitting parameters
for 24 RRab and 2 RRc stars with no apparent signs of amplitude modulations 
(see Fig.~\ref{RRL_A}) are listed in 
Table~\ref{tab:fourier_coeffs}. These Fourier parameters and the above mentioned
calibrations were used in turn to calculate the physical parameters listed in 
Table~\ref{fisicos}. The
absolute magnitude $M_V$ was converted into luminosity with $\log (L/{\rm
L_{\odot}})=-0.4\, (M_V-M_{\rm
bol}+BC$). The bolometric correction was calculated using the formula $BC=
0.06\, {\rm [Fe/H]}_{ZW}+0.06$ given by Sandage \& Cacciari (1990). We
adopted  $M^\odot_{\rm bol}=4.75$~mag. 

Before the iron calibration of Jurcsik \&
Kov\'acs (1996) for RRab stars can be applied to the light curves, a
``compatibility condition parameter'' $D_m$ should be calculated. For the definition
of $D_m$ see the works of
Jurcsik \& Kov\'acs (1996) and  Kov\'acs \& Kanbur (1998). These authors advise to
consider only light curves for which $D_m < 3.0$. The values of $D_m$ for each of the
RR~Lyrae stars are also listed in Table~\ref{tab:fourier_coeffs}. Most of them
fulfill the $D_m$ criterion and, in order to maintain 
the size of our sample reasonable, we did not exclude stars  V1--V14, which have values 
marginally larger than 3.0. This practice has been followed by
some previous authors (e.g. Arellano Ferro et al.\ 2013a; Kains et al.\ 2013, Cacciari
et al.\ 2005).
Stars V25 and V36 show large uncertainties in their Fourier 
coefficients as well as large values of
$D_m$, so they were not considered in the calculation of physical quantities.

\begin{table*}
\scriptsize
\begin{center}
\caption[] {Fourier coefficients of  \RRab and \RRc stars in NGC~3201. 
The numbers in parentheses indicate
the uncertainty on the last decimal place. Also listed are the number of
harmonics~$N$ used to fit the light curve of each variable and the deviation 
parameter~$D_{\textit{\lowercase{m}}}$.}       
\label{tab:fourier_coeffs}   
\begin{tabular}{llllllllllrr}
\hline
Variable     & $A_{0}$    & $A_{1}$   & $A_{2}$   & $A_{3}$   & $A_{4}$   &$\phi_{21}$ & $\phi_{31}$ & $\phi_{41}$ 
& $E(B-V)$ &$N$   &$D_m$ \\
  ID     & ($V$ mag)  & ($V$ mag)  &  ($V$ mag) & ($V$ mag)& ($V$ mag) & & & & & & \\
\hline
\multicolumn{12}{c}{RRab} \\
\hline
V1 & 14.841(2)& 0.331(2)& 0.171(2)& 0.106(2)& 0.065(2)&4.030(19) &8.277(30) & 6.204(45)&0.301&7&5.2\\
V2 & 14.839(2)& 0.357(3)& 0.171(2)& 0.144(2)& 0.083(2)&3.721(18) & 7.906(24)& 5.951(36)&0.247&10&3.1\\
V8 & 14.760(1)& 0.173(1)& 0.085(1)& 0.049(1)& 0.017(1)& 3.962(19)& 8.474(29)& 6.645(72)&0.214&9&7.6\\
V9 & 14.834(2)& 0.318(3)& 0.092(3)& 0.030(3)& 0.016(2)& 3.864(28)& 8.122(46)& 5.872(67)&0.201&7&3.8\\
V13& 14.883(4)& 0.296(2)& 0.148(2)& 0.109(2)& 0.063(2)& 3.953(21)& 8.285(31)& 6.302(43)&0.186&9&4.9 \\
V14& 15.009(3)& 0.400(4)& 0.186(4)& 0.134(5)& 0.095(3)& 3.761(33)& 7.910(47)& 5.824(74)&0.223&9&5.4 \\
V17 &14.820(2)& 0.291(3)& 0.132(3)& 0.095(3)& 0.064(2)& 3.918(26)& 8.262(37)& 6.433(54)&0.224&7&1.4\\
V21& 14.858(1)& 0.267(1)& 0.115(1)& 0.080(1)& 0.047(1)& 3.991(11)& 8.374(17)& 6.484(26)&0.239&7&1.2\\
V23 &14.818(1)& 0.268(2)& 0.132(2)& 0.091(2)& 0.055(2)& 4.012(17)& 8.283(25)& 6.478(37)&0.222&7&1.3\\
V25& 14.787(5)& 0.278(4)& 0.107(5)& 0.062(4)& 0.033(6)& 3.882(55)& 8.179(137)& 6.289(260)&--&10&12.1\\
V35& 14.771(1)& 0.228(1)& 0.104(1)& 0.065(1)& 0.034(1)& 4.068(15)& 8.482(23)& 6.980(39)&0.215&7&1.2\\
V36& 14.751(4)& 0.280(6)& 0.093(6)& 0.037(5)& 0.016(5)& 3.712(52)& 7.773(139)& 5.102(312)&--&8&9.4 \\
V37& 14.785(1)& 0.312(1)& 0.121(1)& 0.067(1)& 0.039(1)& 4.063(9) & 8.353(14)& 6.392(24)&0.228&7&2.2\\
V38& 14.836(1)& 0.303(2)& 0.133(2)& 0.077(2)& 0.039(2)& 3.818(18)& 8.068(30)& 5.963(52)&0.18:&9&1.6 \\
V39& 14.879(2)& 0.430(3)& 0.193(3)& 0.150(3)& 0.101(3)& 3.804(20)& 7.899(27)& 5.839(36)&0.228&9&1.0 \\
V40& 14.854(1)& 0.162(1)& 0.060(1)& 0.033(1)& 0.012(1)& 4.182(19)& 8.611(31)& 7.287(68)&0.235&5&2.7\\
V45& 14.974(3)& 0.372(5)& 0.140(5)& 0.115(5)& 0.081(5)& 3.915(43)& 8.095(57)& 5.952(81)&0.249&10&2.6\\
V49& 14.760(1)& 0.355(2)& 0.139(2)& 0.090(2)& 0.058(2)& 4.049(16)& 8.180(24)& 5.978(36)&0.213&7&2.3\\
V50& 14.832(2)& 0.318(4)& 0.135(3)& 0.104(3)& 0.076(4)& 3.841(34)& 8.018(46)& 6.025(62)&0.18:&10&1.0\\
V76& 14.821(2)& 0.249(3)& 0.112(3)& 0.058(3)& 0.029(3)& 3.964(31)& 8.427(54)& 6.828(80)&0.200&8&1.9\\
V77& 14.693(2)& 0.254(2)& 0.153(2)& 0.097(2)& 0.052(3)& 4.490(23)& 9.115(37)& 7.200(60)&0.19:&8&2.5 \\
V80& 14.650(2)& 0.241(3)& 0.137(3)& 0.101(2)& 0.063(3)& 4.187(27)& 8.539(39)& 6.626(55)&0.17:&8&2.3 \\
V90& 14.753(2)& 0.332(4)& 0.191(4)& 0.139(4)& 0.085(4)& 3.986(30)& 8.153(44)& 5.986(65)&0.213&9&2.5 \\
V100& 14.813(2)& 0.324(3)& 0.152(2)& 0.128(2)& 0.078(2)& 3.911(23)& 7.999(29)& 5.886(43)&0.210&9&1.9 \\
\hline
\multicolumn{12}{c}{RRc} \\
\hline
V11 & 14.802(1)& 0.235(2)& 0.060(2)& 0.027(2)& 0.015(2)&4.589(40) &3.111(77) &
1.891(158)&0.23&4& \\
V98 & 14.759(2)& 0.191(1)& 0.018(1)& 0.010(1)& 0.001(1)&4.826(102) &3.711(187) &
4.653(742)&0.23&4& \\
\hline
\end{tabular}
\end{center}
\end{table*}

\begin{table*}
\footnotesize
\begin{center}
\caption[] {\small Physical parameters of the \RRab and \RRc stars. The
numbers in parentheses indicate the uncertainty on the last 
decimal places and have been calculated as described in the text.}
\label{fisicos}
 \begin{tabular}{lccccccc}
\hline 
Star&[Fe/H]$_{ZW}$ & $M_V$ & log~$T_{\rm eff}$  &log$(L/{\rm L_{\odot}})$ &D (kpc)& 
$M/{\rm M_{\odot}}$&$R/{\rm R_{\odot}}$\\
\hline
\multicolumn{8}{c}{RRab} \\
\hline
V1& $-$1.589(28)&0.523(3)&3.806(9)&1.691(1)&4.75&0.69(7)& 5.66(1)\\
V2& $-$1.665(23)&0.628(4)&3.809(8)&1.649(2)&4.89&0.71(7)& 5.25(1)\\ 
V8& $-$1.487(25)&0.627(1)&3.800(10)&1.649(1)&4.94&0.62(8)& 5.34(1)\\
V9& $-$1.436(43)&0.590(4)&3.814(10)&1.664(2)&5.30&0.71(9)&4.81(2)\\
V13& $-$1.468(29)&0.608(3)&3.808(8)&1.657(1)&5.40&0.66(7)&5.064(1)\\
V14& $-$1.572(44)&0.607(6)&3.815(11)&1.657(2)&5.52&0.73(9)&5.115(2)\\ 
V17& $-$1.452(35)&0.616(4)&3.808(9)&1.653(2)&5.03&0.67(7)&5.01(1)\\
V21& $-$1.348(16)&0.631(1)&3.809(7)&1.647(1)&4.98&0.65(6)&5.02(1)\\
V23& $-$1.511(23)&0.610(3)&3.805(8)&1.656(1)&5.06&0.66(6)&5.12(1)\\
V35& $-$1.435(22)&0.595(1)&3.801(8)&1.662(1)&5.03&0.66(6)&5.27(1)\\
V37& $-$1.419(13)&0.549(1)&3.809(7)&1.680(1)&5.08&0.69(6)&5.09(1)\\
V38& $-$1.425(28)&0.672(3)&3.814(9)&1.631(1)&4.93&0.68(7)&4.72(2)\\
V39& $-$1.484(25)&0.628(4)&3.818(8)&1.649(2)&5.11&0.73(7)&4.58(1)\\
V40& $-$1.422(29)&0.608(1)&3.796(10)&1.657(1)&5.05&0.64(8)&5.41(1)\\
V45& $-$1.504(54)&0.580(7)&3.813(11)&1.668(3)&5.30&0.70(9)&4.87(2)\\
V49& $-$1.595(23)&0.513(3)&3.809(8)&1.695(1)&5.22&0.71(7)&5.53(1)\\
V50& $-$1.594(43)&0.627(5)&3.809(10)&1.649(2)&5.02&0.69(8)&5.30(2)\\
V76& $-$1.153(51)$^{a}$&0.692(4)&3.812(11)&1.623(2)&5.03&0.65(9)&4.81(2)\\
V77& $-$0.660(35)$^{a}$&0.657(3)&3.820(10)&1.637(1)&4.65&0.55(6)&5.03(1)\\
V80& $-$1.282(37)$^{a}$&0.646(4)&3.808(9)&1.642(2)& 4.58&0.61(7)&5.14(2)\\
V90& $-$1.711(41)&0.547(6)&3.805(10)&1.681(2)&5.12&0.68(8)&5.66(2)\\
V100& $-$1.638(27)&0.629(4)&3.809(8)&1.648(2)&5.09&0.68(7)&5.34(1)\\
\hline
Weighted mean& $-$1.483(6)&0.604(1)&3.808(2)&1.658(1)&5.00&0.67(2)&5.12(1)\\
$\sigma$&$\pm$0.098&$\pm$0.045&$\pm$0.005&$\pm$0.016&$\pm$0.22&$\pm$0.03&$\pm$0.25\\
\hline
\multicolumn{8}{c}{RRc} \\
\hline
V11& $-$1.46(14)&0.574(2)&3.868(1)&1.670(1)&5.04&0.57(1)&4.15(4)\\
V98& $-$1.55(37)&0.590(4)&3.864(1)&1.664(2)&4.91&0.50(1)&4.78(12)\\
\hline
Weighted mean& $-$1.47(12)&0.576(1)&3.867(1)&1.667(2)&5.03&0.56(1)&4.21(4)\\
$\sigma$&$\pm$0.06&$\pm$0.011&$\pm$0.003&$\pm$0.016&$\pm$0.09&$\pm$0.05&$\pm$0.45\\
\hline
\end{tabular}
\end{center}
\raggedright
\center{\quad $^{a}$Value not considered in the average.}
\end{table*}

The resulting physical parameters of the RR~Lyrae stars are
summarized in Table~\ref{fisicos}. The mean values given in the bottom of the table
are weighted by the statistical uncertainties. Also listed are the corresponding
distances.
Given the period, luminosity, and temperature for each RR~Lyrae star, its
mass and radius can be estimated from the equations: $\log~M/M_{\odot} =
16.907 - 1.47~ \log~P_F + 1.24~\log~(L/L_{\odot}) - 5.12~\log~T_{\rm eff}$ (van Albada
\& Baker 1971), and $L$=$4\pi R^2 \sigma T^4$ respectively.
The masses and radii given in Table~\ref{fisicos} are expressed in solar
units. 

\subsection{Bailey diagram and Oosterhoff type}
\label{sec:Bailey}
\begin{figure} 
\includegraphics[width=8.cm,height=13.cm]{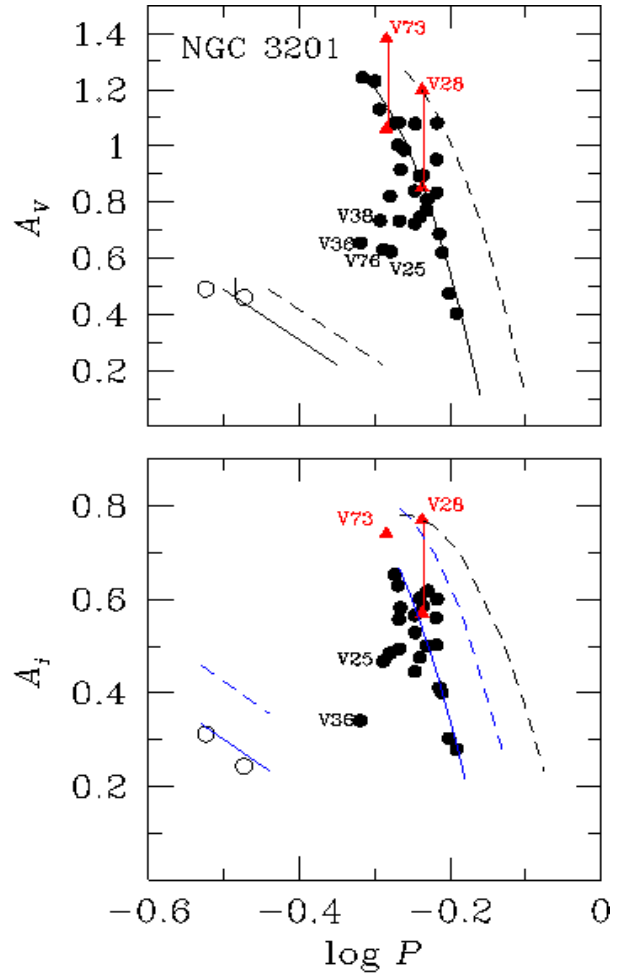}
\caption{Bailey diagram of NGC~3201 for $V$ and $I$ amplitudes. In the
top panel the continuous and segmented lines are the loci found by Cacciari et al.\
(2005) in the OoI cluster M3. In the bottom panel the black segmented locus was found
by
Arellano Ferro et al.\ (2011; 2013a) for the OoII clusters NGC~5024 and NGC~6333. The
blue loci are from Kunder et al. (2013). See \S~\ref{sec:Bailey} for details.}
    \label{Fig:Bailey}
\end{figure}

The Bailey diagram is a plot of the period versus the amplitude for RR~Lyrae stars;
it offers insight on the Oosterhoff type of a globular cluster and helps to
identify
possible peculiar amplitude stars for a given period. The Bailey diagram of
M3 is usually used as a reference for OoI~clusters (see Fig.~4 of Cacciari et al.\
2005).
Fig.~\ref{Fig:Bailey} displays the corresponding distribution of the RR~Lyrae stars
in NGC~3201 with a good light curve coverage. The continuous and segmented
lines in
the top diagram represent, respectively, the mean distributions of non-evolved and
evolved stars in M3 according to Cacciari et al.\ (2005). In recent papers we compared
these loci with the distributions of RR~Lyrae stars in the OoII clusters NGC~5024
(Arellano
Ferro et al.\ 2011, Fig.~7), NGC~6333 (Arellano Ferro et al.\ 2013a, Fig.~17), and NGC
7099 (Kains et al. 2013, Fig. 10). 
We, like Cacciari et al.\ (2005),  found that the $A_V$
amplitudes of RRab stars follow the loci of the candidate evolved stars in M3, i.e.,
the segmented line; as for the RRc stars, however, we differ from Cacciari et al.\
(2005) in that the distributions in the OoII clusters NGC~5024 and NGC~6333 do
not follow the defined trend in M3~RRc stars but are rather scattered, and that their
distribution is altered by the presence of Blazhko-like amplitude modulations,
particularly in the case of NGC~5024 (Arellano Ferro et al.\ 2012).
It is clear from this figure that the RRab and
RRc stars in NGC~3201 follow the trend found in M3, which identifies NGC~3201 as
being of the type OoI. In Arellano Ferro et al.\ (2011) the distribution of $A_I$
amplitudes in the
OoII cluster NGC~5024 was also defined and is shown as a black segmented line in the
bottom
panel of Fig.~\ref{Fig:Bailey}. It has the eqution:

\begin{eqnarray}\label{eq:AIOoIIab}
A_I &=& (-0.313 \pm 0.112) - (8.467 \pm 1.193)\, \log P \nonumber \\
 &&- (16.404 \pm 0.441)\, \log P^2.
\end{eqnarray}

Also in the bottom panel of Fig.~\ref{Fig:Bailey} the distribution of the $I$ 
amplitudes of RRab and RRc stars in NGC~3201 is shown.
The blue locus are those calculated by Kunder et al. (2013)
for the OoI clusters for the RRab and RRc stars (solid lines) and for the OoII
clusters (segmented lines). Their OoI locus represents well the
distribution in NGC 3201. We note the difference in the locus of OoII
clusters proposed by Kunder et al. (2013) and the one observed by Arellano Ferro et
al. (2011; 2013a) in NGC~5402 and NGC~6333 respectively.

\section{SX Phe stars: The Period-Luminosity  relation}
\label{sec:SXP}

The Period-Luminosity relation for SX Phe stars (PLSX) has recently been calibrated 
 by several
authors, notably Poretti et al.\ (2008) and McNamara (1997) for Galactic and
extragalactic $\delta$~Scuti and SX~Phe stars. In globular clusters the PLSX 
has been studied by Jeon et al.\
(2003) and Arellano Ferro et al.\
(2011) for NGC~5024, and McNamara (2000) for $\omega$~Cen. 
The calibrations of Arellano Ferro et al.\ (2011) for the fundamental mode of SX Phe
stars in NGC~5024 in the \textit{V} and \textit{I} filters are of the form:
\begin{equation}
\label{PLV}
M_V = -2.916\, \log P - 0.898,
\end{equation}
and
\begin{equation}
  \label{PLI}
M_I = -2.892\, \log P - 1.072.
\end{equation}
\noindent
These  calibrations have been used to calculate the distance to
globular clusters with SX~Phe stars independently of their RR~Lyrae population 
(e.g., Arellano Ferro et al.\ 2013a;b).

\begin{figure} 
\includegraphics[width=8.cm,height=13.cm]{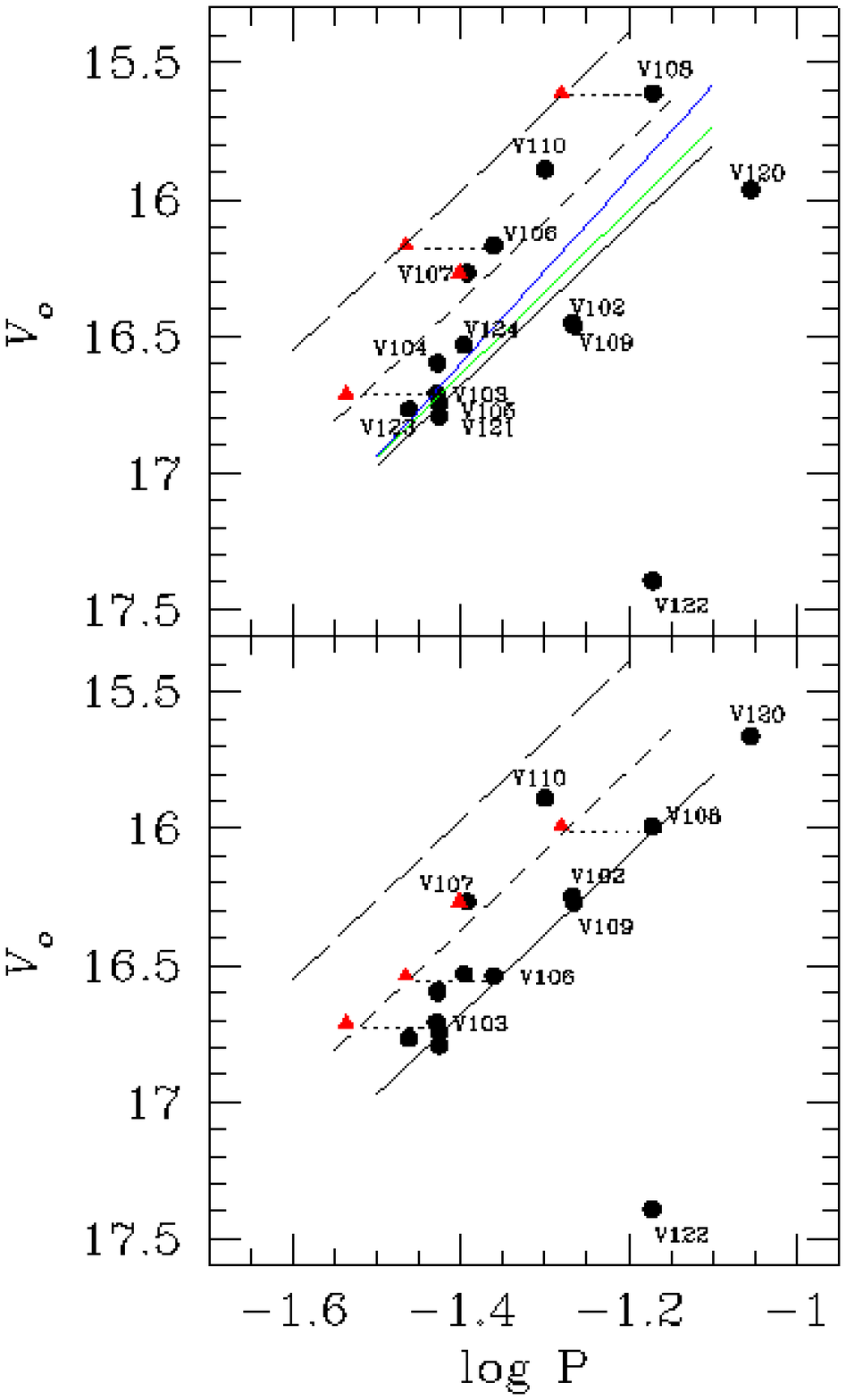}
\caption{Period-Luminosity (P-L) relation for SX~Phe stars. 
In the top panel the black lines correspond to
the fundamental mode P-L
calibration of Arellano Ferro et al.\ (2011) for the SX~Phe in NGC~5024, and their
corresponding first and second overtone loci, all
shifted to the distance of 5.0~kpc obtained from the RR~Lyrae stars. 
The blue and green lines are the fundamental mode calibrations of Cohen \& Sarajedini
(2012) and Jeon et al. (2003). 
All these loci were also shifted to the distance of 5.0~kpc.
Magnitudes and loci have been dereddened by $E(B-V)=0.23$. The scatter shown by the
star distribution is mostly
due to  differential reddening. In the bottom panel we modified the individual
reddenings to reconcile the double-mode pulsators V108 and V109 with the fundamental 
and first overtone loci. Also
small shifts were applied to V102, V109, and V120. Stars V107 and V110 were not
moved. 
Star V107 has an active non-radial mode, while the position of V110 may indicate
that it
pulsates either in the first overtone or that its reddening is smaller. Variable V122 is
likely
not a cluster member. Stars V124, V104, V105, V123, and V121, in order of
increasing brightness, are not labeled in the bottom panel. See $\S$~\ref{sec:SXP} for
a discussion.}
    \label{Fig:PLSX}
\end{figure}

In the top panel of Fig.~\ref{Fig:PLSX} the distribution of the SX~Phe stars is shown
in  the $\log P - V_0$
plane. To deredden them we adopted the mean reddening estimated 
for the RR Lyrae stars in \S~\ref{sec:RRL},  $E(B-V)=0.23$. The continuous line
corresponds to the calibration for
the fundamental mode of Eq.~\ref{PLV} scaled to the distance of 5.0~kpc, obtained
from the 22~RRab stars in Table~\ref{fisicos} (see \S~\ref{sec:Four}).

The dotted and dashed lines correspond to the locus of the first and second overtones
assuming the ratios 0.783 and 0.571 (see Santolamazza et al.\ 2001 or Jeon et al.\
2003; Poretti et al.\ 2005). It is clear that the distribution of SX~Phe stars shows
scatter, which is most likely due to  differential reddening. However, we note that
the group around V103 matches the fundamental mode rather well.
 It is also worth paying attention to the position of the three double-mode stars
V103, V106, and V108, for which we also plotted the first overtone period with
red triangles. In these stars we detected the fundamental and first overtone
frequencies, identified as such from their ratio (\S~\ref{Sec:SXPHE}).
To reconcile the positions of the stars with the pulsation loci, one may assume a
slightly
different reddening for each star. In this way we obtained the bottom panel of
Fig.~\ref{Fig:PLSX}, and the
 procedure was as follows. The two double-mode stars V106 and V108 were shifted to
the fundamental mode line by
modifying their reddening  to $E(B-V)=0.11$ in both cases. Note
that the corresponding first overtone mode, represented by the red triangles, also
matches well the first overtone locus for these double-mode stars. Small reddening
adjustments were also applied to stars
V102, V109, and V120. The estimated reddening values for all SX~Phe stars are
listed in the 4th column of Table~\ref{SXfreq}. Star V122 is much too faint for
its period, and
an unacceptable large value of the reddening would be required to bring it to the
fundamental locus. For these reasons, we believe that this SX~Phe star is not a
cluster member.

We note at this point that the slopes of the $V$ and $I$ P--L relations for NGC 3201
seem consistent with those observed in NGC 5024 (Arellano Ferro et al.\ 2011) and
also with the $V$ slopes observed in NGC 5024 by Jeon et al.\ (2003) (green line
in Fig. \ref{Fig:PLSX}) and in NGC~288 (Arellano Ferro et al.\ 2013b). The slope found
by Cohen \& Sarajedini (2012) for the fundamental mode is a little steeper (blue line
in Fig. \ref{Fig:PLSX}). Both Jeon et al.'s\ (2003) and Cohen \& Sarajedini's\
(2012) calibrations are
consistent with our distance determination of 5.0 kpc; however, if these calibrations
are preferred, the individual reddenings of the SX Phe discussed above would have to
be
slightly modified.

\begin{table*}
\footnotesize
\caption{Time-series \textit{V} and \textit{I} photometry for all the confirmed variables in our 
field of view. The standard \Mstd and
instrumental \mins magnitudes are listed in columns 4 and~5,
respectively, corresponding to the variable stars in column~1. Filter and epoch of
mid-exposure are listed in columns 2 and 3, respectively. The uncertainty on
\mins is listed in column~6, which also corresponds to the
uncertainty on \Mstd. For completeness, we also list the
reference and differential fluxes \fref 
and \fdif and the scale factor \lowercase{\textit{p}} 
in columns 7, 9, and~11, along with the uncertainties \sref 
and \sdiff in columns 8 and~10. This is an extract from
the full table, which is available with the electronic version of the article.
         }
\centering
\begin{tabular}{ccccccccccc}
\hline
Variable &Filter & HJD & $M_{\mbox{\scriptsize std}}$ &
$m_{\mbox{\scriptsize ins}}$
& $\sigma_{m}$ & $f_{\mbox{\scriptsize ref}}$ & $\sigma_{\mbox{\scriptsize ref}}$ &
$f_{\mbox{\scriptsize diff}}$ &
$\sigma_{\mbox{\scriptsize diff}}$ & $p$ \\
Star ID  &    & (d) & (mag)     & (mag)   & (mag) & (ADU s$^{-1}$) &(ADU s$^{-1}$)   
               &(ADU s$^{-1}$)  &(ADU s$^{-1}$)    & \\
\hline
V1 &$V$ &2456371.52286& 15.126& 16.348 & 0.001 &  6014.959 &  4.635 & $-$3115.105 & 
3.956 & 0.9966\\
V1 &$V$ &2456371.52599& 15.137& 16.359 & 0.001 &  6014.959 &  4.635 &
$-$3157.769 & 3.334 & 1.0006\\
\vdots   &  \vdots  & \vdots & \vdots & \vdots & \vdots   & \vdots & \vdots  & \vdots&
\vdots \\
V1 &$I$ &2456371.51722&  14.053&  15.830&  0.002&   6831.708&   7.004&  $-$2176.444&  
9.449 & 1.0030
\\
V1 &$I$ &2456371.51972&  14.017&  15.793&  0.002&   6831.708 &  7.004 & $-$2018.446& 
9.645&  1.0047\\
\vdots   &  \vdots  & \vdots & \vdots & \vdots & \vdots   & \vdots & \vdots  & \vdots&
\vdots \\
 V2&$V$& 2456371.52286&  14.457&  15.685&  0.001&   2795.928&   4.714&  $+$2514.549 & 
4.843 & 0.996\\
 V2 &$V$&2456371.52598 & 14.477&  15.706&  0.001&   2795.928&   4.714 & $+$2424.449 &
3.992 & 1.0006\\
\vdots   &  \vdots  & \vdots & \vdots & \vdots & \vdots   & \vdots & \vdots  & \vdots&
\vdots \\
 V2&$I$& 2456371.51722&  13.687&  15.466&  0.002&   4060.533&   7.058&  $+$2464.694 &
11.144&  1.0030\\
 V2 &$I$&2456371.51972&  13.799&  15.578&  0.002&   4060.533&   7.058&  $+$1827.676 &
11.074&  1.0047\\
\vdots   &  \vdots  & \vdots & \vdots & \vdots & \vdots   & \vdots & \vdots  & \vdots&
\vdots \\
\hline
\end{tabular}
\label{tab:vi_phot}
\end{table*}

We may conclude that the SX~Phe stars indicate a distance to the cluster
consistent with that derived by the Fourier decomposition of the RRab
stars, i.e., 5.0~kpc, and that by invoking adequate values
for the differential reddening of these stars it can be argued that their main
frequency corresponds to the fundamental mode, and that V103, V106, and V108 are
double-mode radial pulsators.

\section{Summary of results}
\label{sec:Summ}

We have presented the results of the Fourier decomposition of RR~Lyrae stars
and the frequency analysis of SX~Phe stars in the central parts of the globular
cluster NGC~3201, based on difference image analysis of a CCD time-series.

The differential reddening of the cluster was addressed and
individual reddenings for the RRab stars were estimated from $V-I$ curves using
the method outlined and calibrated most recently by Guldenschuh et al.\ (2005). We
found an average $E(B-V)=0.23\pm0.02$ from a sample of 18 RRab stars with adequate phase
coverage. In the calculation of the distance, however, we used
the individual reddenings.

Iron abundance and distance were calculated from the light curve Fourier decomposition
of 22 RRab and 2 RRc stars contained in the \fov of our images without obvious signs
of amplitude modulation. We found, for the RRab stars, a
mean iron abundance [Fe/H]$_{ZW}=-1.483\pm0.090$ (systematical uncertainty) and a mean
distance of $5.00\pm0.22$~kpc (systematical).
For the RRc stars the
results were, respectively: [Fe/H]$_{ZW}=-1.47\pm0.06$ and 
 $5.03\pm0.09$~kpc (both are also systematical uncertainties). Since these results
come
from independent calibrations for the RRab and RRc stars, they can be considered as 
two independent determinations of metallicity and distance.
The iron abundance in the scale of Carretta et al.\ (2009), transformed using the
equation:
%
%
\begin{eqnarray}\label{UVES}
[\mathrm{Fe/H}]_{\mathrm{UVES}} &=& -0.413 +0.130\, [\mathrm{Fe/H}]_{ZW} \nonumber \\
  && -0.356\, [\mathrm{Fe/H}]_{ZW}^2,
\end{eqnarray}
%
is [Fe/H]$_{\mathrm{UVES}}=-1.39 \pm 0.13$. To the best of our knowledge no previous estimates
of [Fe/H] from Fourier decomposition of the RR~Lyrae light curves exist for this
cluster. Other estimates of [Fe/H] include: $-1.54 \pm 0.16$ and $-1.89 \pm 0.16$
calculated by LS03 from their $B-V$ and $V-I$ CMDs, $-1.53 \pm 0.03$ (Rutledge et al.\
1997), and $-1.61 \pm 0.12$ (Zinn \& West 1984). Within the uncertainties, our metallicity
[Fe/H]$_{ZW}=-1.483\pm0.090$ is in good concordance with these published values.

The calculation of the cluster distance by LS03 was made via the globular cluster
$M_V$ and [Fe/H] correlation (e.g., Chaboyer 1999) for RR~Lyrae stars, and yielded
$4.87\pm0.27$~kpc for [Fe/H]$=-1.53$. This is in good agreement with the result of our RR~Lyrae
Fourier
decomposition. LS03 also calculated the cluster distance based on two SX~Phe
stars, \#752 and \#1019 (in their numeration) which, judging from their mean magnitudes and
periods, very likely correspond to variables V110 and V109, respectively. They used the PLSX
calibration of Petersen \& Hog (1998) and derived a distance of $4.67\pm0.24$~kpc. Our
own calculation
is based on a larger number of SX~Phe, a more complete mode identification for most of
them, some assumptions on the differential reddening, and the use of a different P-L
calibration (see \S~\ref{sec:SXP}). It was shown that, in the P-L distribution of
SX~Phe stars,
both the fundamental and the first overtone modes loci are consistent with the
average distance $5.00\pm0.23$~kpc found from the RR~Lyrae stars. 

Finally, we discovered  three new SX~Phe stars, numbered V122, V123, and V124. Very likely, star V122 is
not a cluster member but a background object. Three clear radial double-mode SX~Phe
stars have been identified, i.e., V103, V106, and V108. Star V107 seems to be a double-mode star
with a non-radial component.

\vskip 2.0cm

{\bf APENDIX A}

Our $V$ and $I$ photometry for all variable stars in the FoV of our images is only
available in electronic form. Here, in Table~\ref{tab:vi_phot}, we show only a
portion of it for guidance regarding its form and content.

\vskip 1.0cm
\noindent
We warmly thank Dr.~Daniel Bramich for allowing us the use DanDIA and 
for guiding our reduction process, as well as for useful and opportune comments.
We are thankful to our referee Christine Clement for her constructive
suggestions and comments.
AAF acknowledges Roberto Figuera Jaimes for his valuable help in the first
stages of the data reduction.
We are indebted to the CONACyT (M\'exico) and MINCyT (Argentina) for financial
support through the interchange project 188769 (MX/12/09). AAF acknowledges the 
support from DGAPA-UNAM grant through project IN104612, to the European Southern
Observatory (Garching), and to the Observatorio Astron\'omico of the Universidad Nacional de
C\'ordoba (Argentina) for warm hospitality during the different stages of this work.
We have made an extensive use of the SIMBAD and ADS services, for which we are
thankful.

\end{document}